# Development of Wire + Arc Additive Manufacturing for the production of large-scale unalloyed tungsten components

G. Marinelli*, a, F. Martina a, S. Ganguly a, S. Williams a

aWelding Engineering and Laser Processing Centre (WELPC), College Road, Cranfield University, Cranfield, MK43 0AL, UK

*Corresponding Author. E-mail address: g.marinelli@cranfield.ac.uk (G. Marinelli)

## Abstract

The manufacturing of refractory-metals components presents some limitations induced by the materials' characteristic low-temperature brittleness and high susceptibility to oxidation. Powder metallurgy is typically the manufacturing process of choice. Recently, Wire + Arc Additive Manufacturing has proven capable to produce fully-dense large-scale metal parts at relatively low cost, by using high-quality wire as feedstock. In this study, this technique has been used for the production of large-scale tungsten linear structures. The orientation of the wire feeding has been studied and optimised to obtain defect-free tungsten deposits. In particular, front wire feeding eliminated the occurrence of pores and micro-cracks, when compared to side wire feeding. The microstructure, the occurrence of defects and their relationship with the deposition process have also been discussed. Despite the repetitive thermal cycles and the inherent brittleness of the material, the as-deposited structures were free from internal cracks and the layer dimensions were stable during the entire deposition process. This enabled the production of a relatively large-scale component, with the dimension of 210 x 75 x 12 mm. This study has demonstrated that Wire + Arc Additive Manufacture can be used to produce large-scale parts in unalloyed tungsten by complete fusion, presenting a potential alternative to the powder metallurgy manufacturing route.

## 1 Introduction

For quite some time, the research within the energy sector has been focused on the development of alternative sources, such as nuclear fusion [1]. In this environment, material selection is key to ensure an optimum service life for the structural components of the reactor. Materials with excellent mechanical properties, heat resistance and neutron load-capacity are required to produce high-quality components.

Refractory metals and their alloys represent one of the most suitable classes of materials for nuclear fusion reactors. Among these, tungsten is one of the candidates for high performance and advanced components in the field of innovative high-temperature energy generation and conversion systems [1,2]. In particular, tungsten can be used as plasma facing material in future fusion reactors [1,2], thanks to its attractive physical properties. In fact, it is characterised by the highest melting point amongst metals, high density, relatively high thermal conductivity, low coefficient of thermal expansion, excellent mechanical properties at elevated temperatures, high resistance to sputtering and erosion, and low tritium retention [3,4].

Owing to the extremely high melting point, tungsten is generally manufactured via the powder metallurgy route [5]. The combination of sintering and hot or cold-work plays a fundamental role in powder consolidation and design of the final structure. However, the low fracture toughness, the high ductile-to-brittle-transition temperature (DBTT) and the high recrystallization temperature present significant challenges for the manufacturing of tungsten components [6,7].

Currently, there are three main manufacturing operations that are being studied, when referring to tungsten components for nuclear fusion environment: the industrial production of large-scale components; the joining of these parts with other materials; and their efficient repair and maintenance.

Additive Manufacturing (AM) can be one of the most auspicious manufacturing technologies for such an advanced alloy [8,9]. Three-dimensional structures can be promptly deposited starting from 3D-models, using a layer-by-layer approach [10]. The practical advantages are cost reduction, freedom of design and control of mechanical properties [10]. Numerous researches have already reported the potential of AM techniques for metallic materials, some of them are summarised in the review of Frazier [11] and Herderick [12].

AM could definitely address some of the manufacturing issues related to tungsten components, and possibly enable the development of new designs approaches. In particular, tungsten tiles for the nuclear fusion divertor can be produced in a more efficient way; their properties could be fine-tuned according to the final application. Additionally, functionally-graded structures produced with AM could lead to controlled and tailored composition and properties, which can enhance the structural integrity and overall performances of the components. Finally, due to the modularity and the extreme flexibility of some AM techniques, f.i. robot-based manipulation, the repair and the maintenance of tungsten plasma-facing components could be accomplished directly in-situ. This would markedly decrease nuclear fusion reactors' operating costs.

In the past, it has been reported that tungsten exhibits limited weldability, due to the low DBTT, the high melting point and the high reactivity with environmental gasses over a large range of temperature [13–16]. However, in a previous study [17], we have been able to weld tungsten autogenously, avoiding cracks and porosity successfully.

With regards to AM, laser and powder have been adopted in various studies, as the heat source and the feedstock, respectively [18]. However, to the authors' best knowledge, most of the investigations available in the literature were unable to achieve fully-dense deposits, and the predominant occurrence of micro-cracks and pores was not avoided [19–22]. These issues must be addressed in order to achieve the structural integrity required for nuclear application, critically to avoid failures and also increase the lifetime under neutron radiation.

In the feasibility study of Nie et al. [19], tungsten cubes were produced using laser-powder-bed-fusion (LPBF), using a femtosecond laser. The components were affected by discontinuities and porosity throughout their volume. Wang et al. [20] also produced pure tungsten components using LPBF, achieving densities as high as 96% of tungsten's theoretical one. Their study showed that when using polyhedral tungsten powder there was a marked occurrence of balling, resulting in large defects and pores; instead, when using a spherical powder, continuous tracks were achieved although a dense network of micro-cracks and porosity were present. Furthermore, Iveković et al. [21] also observed extensive cracking in LPBF parts, even when employing pre-heating of the substrate to 400℃. This was attributed to thermal stress build-up during solidification, which is usually rapid in LPBF, and to the presence of interstitial contaminants such as oxygen, nitrogen or carbon. LPBF solidification dynamics were studied by Zhou et al. [22]. They found that tungsten high surface tension and viscosity induced a relative slow wetting and spreading speeds, enhancing the balling phenomenon.

Differently from LPBF, wire-feed technologies are emerging mainly because of their ability to produce large-scale components [23,24]. Wire + Arc additive Manufacture (WAAM) is a novel AM process which employs an electric arc as the heat source, and high-quality metal wire as the feedstock [25]. WAAM can directly fabricate fully-dense metallic large 3-D near-net-shape components with a much higher deposition rate, than most other metal additive manufacture processes [23,25], the highest rate so far being of 9.5 kg/h [26]. The WAAM process has successfully produced large-scale parts in stainless steel [27], Inconel ® [28], titanium [29] and aluminium [30]. The manufacture of large and engineered components by WAAM is attractive also because of the low system and operating costs, as well as the modularity of the system design [25,31].

The main aim of this research is to apply the WAAM process to unalloyed tungsten, for the first time, with particular attention to achieving defect-free fully-dense tungsten components, on a large scale. The study and monitoring of the metal transfer, and the characterisation from the microstructural point of view are discussed. A structure of realistic scale has also been produced to understand the issues related to scaling up, and ultimately assess the feasibility of WAAM's implementation as an innovative way to produce unalloyed tungsten parts.

## 2 Material and methods

Unalloyed tungsten wire with a diameter of 1 mm and unalloyed tungsten plates produced by powder metallurgy were used as feedstock and substrates, respectively. The plates used for the study on the feeding direction had dimensions of 80 mm in length, 10 mm in width and 30 mm in height, while the plate used for the deposition of the large tungsten linear structure had dimensions of 210 mm in length, 10 mm in width and 30 mm in height. The surface of the plates was ground and rinsed with acetone prior to deposition in order to remove most of the contaminants. The composition of the materials used has been reported in **Table 1**.

**Table 1**
Elemental composition (wt.%) of tungsten for the substrate and the wire used in this study.

|  | W | Mo | Ta | Ti | V | Cr | Fe | C | N | O | K |
|---|---|---|---|---|---|---|---|---|---|---|---|
| **Substrate** | 99.99 | <0.05 | <0.05 | <0.05 | <0.05 | <0.05 | <0.05 | <10 ppm | <10 ppm | <50 ppm | <10 ppm |
| **Wire** | 99.99 | <0.05 | <0.05 | <0.05 | <0.05 | <0.05 | <0.05 | 33 ppm | <10 ppm | <50 ppm | <10 ppm |

**Fig. 1** shows the clamping system used for the deposition. The selection of the substrate orientation with respect to the heat source and the utilisation of a specific design for the clamping system represented important aspects of the process development. Initial trials performed on substrates held in the flat position showed consistent cracking. This was caused by the process-induced thermal shock and by the built up of the residual stresses. Therefore, an unconventional approach was adopted i.e. the deposition was performed on the edge of the substrate, as shown in **Fig. 1.** This prevented the development of lateral cracks. The substrate was firmly held in position by two side-clamping bars with a constant force of 25 N applied using a series of screws. These parts were connected to the main backing support, which was fixed to two lateral supports and to the motorised stages.

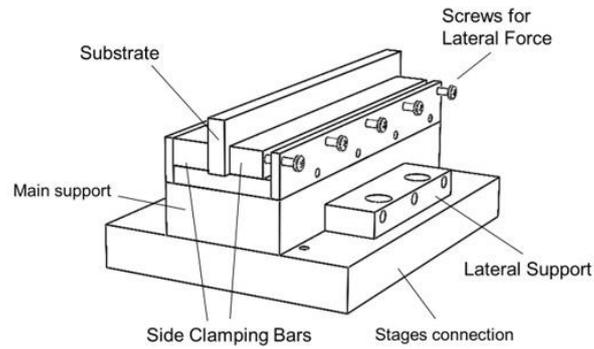

**Fig. 1**: Schematisation of all the components used for the clamping system.

**Fig. 2** shows the apparatus used for the deposition. The main coordinates are kept constant for both the schematics reported. In particular, **Fig. 2a** shows the configuration which employs side wire feeding and **Fig. 2b** shows the configuration for the front wire feeding approach. For both configurations, the layers were progressively deposited onto the substrate using a single bead. The direction of deposition was always kept constant for each successive layer.

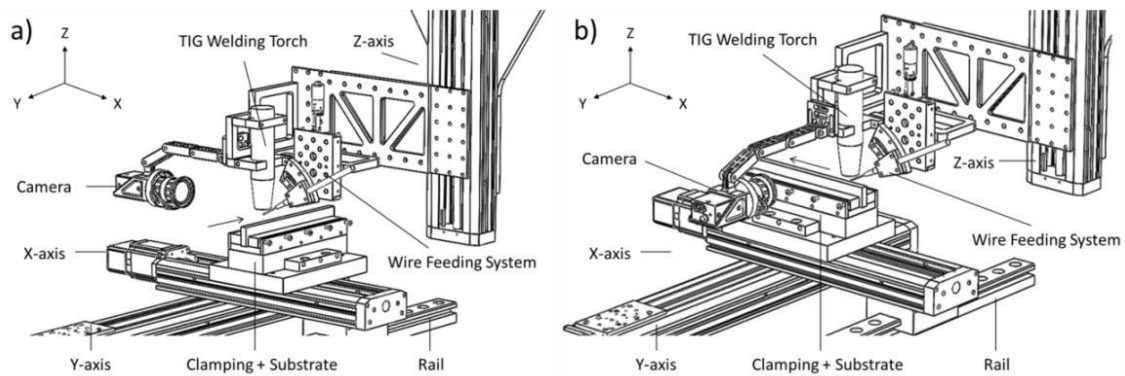

**Fig. 2**: Set-up used for the development of the WAAM process for unalloyed tungsten when using side wire feeding (a) and when using front wire feeding (b).

A conventional tungsten inert gas (TIG) welding torch, a power supply and a controlled wire feeder were used for the deposition. The heat source, the wire delivery system and the substrate were attached to three linear motorized high-load stages assembled in XYZ configuration (**Fig. 2**). The orientation of the camera was always kept perpendicular with respect to the deposition direction. The entire apparatus was contained within an air-tight enclosure, to achieve a $O_2$ concentration of around 100 ppm.

**Table 2** reports the deposition parameters used to produce all the samples analysed in this study. Please note the choice of 100% He as process-gas for TIG, following from what concluded in our previous study [17].

**Table 2**
WAAM process parameters used for the deposition of unalloyed tungsten.

| | |
|---|---|
| **Travel Speed (TS) [mm/s]** | 2 |
| **Welding Current (I) [A]** | 400 |
| **Wire Feed Speed (WFS) [mm/s]** | 35 |
| **Shielding Gas Composition (SGC) [%]** | 100 He |
| **Gas Flow Rate (GFR) [L/min]** | 15 |

Microstructure and defects were examined using the cross-section perpendicular to the deposition direction, for both side-fed and front-fed samples. Cross-sections were ground and polished using paper discs of silicon carbide, and then etched using the Murakami's reagent.

After optimal parameters had been found, a large-scale tungsten wall was built using these. The linear structure produced was characterised by the dimensions of 120 mm in length, 75 mm in height and around 12 mm in thickness.

## 3  Results and Discussion

### 3.1  *Metal transfer from the wire to the weld pool*

#### 3.1.1  Side wire-feeding

**Fig. 3** shows a sequence of time-resolved images of the deposition process when using the side wire feeding configuration. It is clear that the process was characterised by the marked presence of spattered particles. In particular, this was caused by two main ejection mechanisms. In **Fig. 3a**, an ejection stream which was coming directly from the weld pool can be seen. Contrarily, in **Fig. 3c**, the spattered particles seemed to be ejected from the liquid metal at the end of the wire. Spattered particles were already seen for tungsten tiles exposed to helium plasma. Coenen et al. [32] suggested that tungsten erosion may occur during high-temperature exposures by evaporation and macroscopic mass losses from the melt, such as spraying or splashing.

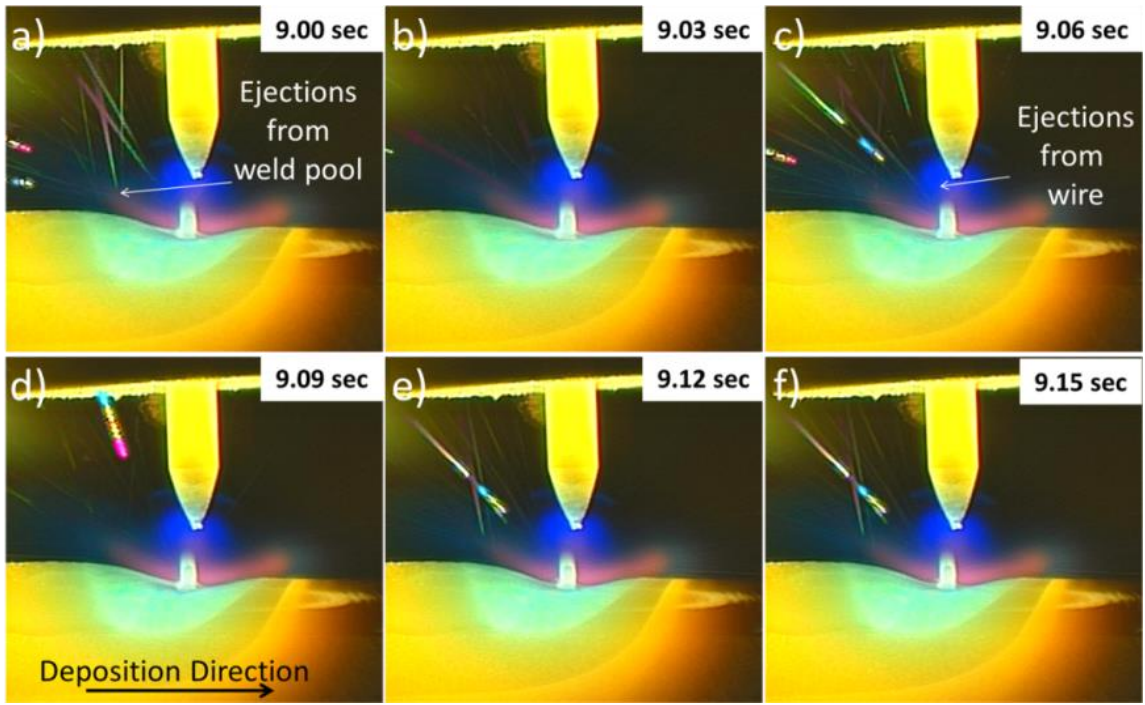

**Fig. 3:** Time-resolved images of the deposition performed using side wire feeding configuration.

The droplets or macroscopic mass losses, localised predominantly along the melted surface, arose because of the Kelvin–Helmholtz instability [32–34]. This phenomenon occurs when there is a difference in velocity across the interface between two fluids. Furthermore, it has been reported that the Kelvin–Helmholtz instability can also lead to the evolution of shock waves along the surface of the fluid causing a breakup of the melt surface into droplets [35].

In WAAM deposition, the Kelvin–Helmholtz instability occurred because of the different relative speeds and different densities of the liquid tungsten and the helium plasma, which created a velocity shear at their interface. In fact, at the centre of the arc column, the plasma gas is characterised by a high velocity [36]. This is more pronounced when using helium instead of argon as shielding gas due to the higher gas velocity at the core of the arc [37]. Thus, the particle ejections occur when liquid tungsten at the tip of the wire faces the high-velocity stream of plasma gasses near the centre of the arc. Furthermore, the particle ejections deriving directly from the weld pool could have been caused by a combined effect of the Kevin-Helmotz instability on the upper surface of the weld pool and the high level of disturbance on the weld pool internal flow caused by the wire itself. This led shock waves to develop within the liquid pool. Their impact could have been high enough to overcome the surface tension, resulting in the ejection of particles from the back of the weld pool.

In **Fig. 4**, the linear structure deposited using side wire feeding is presented. The silver surface suggests a complete absence of oxidation. The two sides of the structure are compared in **Fig. 4b** and **Fig. 4c**.

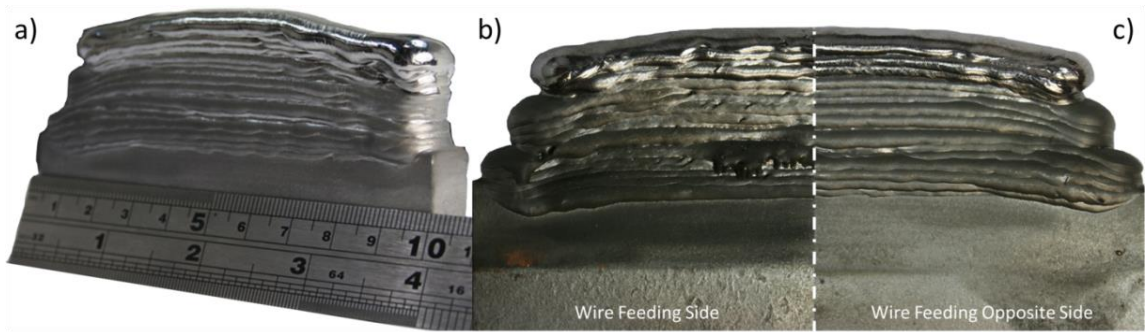

**Fig. 4:** Tungsten linear structure deposited using side wire feeding (a). Comparison of the wire feeding side (b) of the structure and the opposite side (c).

**Fig. 4b** shows the wall appearance of the side from which the wire was delivered to the weld pool, while **Fig. 4c** shows the wall appearance of the opposite side. A large number of irregularities and small defects were found on the outer surface of the wire feeding side (**Fig. 4b**). The occurrence of spattering clearly enhanced the formation of defects like voids and lacks of fusion. Further investigations about the nature and the formation of these defects are reported later within the microstructural analysis section.

### 3.1.2 Front wire-feeding

**Fig. 5** reports a sequence of time-resolved images of the deposition process employing the front wire feeding. There was no presence of spatter or weld pool perturbation during the deposition. The wire was smoothly melted and transferred at a controlled rate to the weld pool by keeping a constant liquid bridge transfer. The ejections from the liquid tungsten at the end of the wire were considerably reduced when the front-fed approach was used. This occurred because the wire was promptly transferred to the liquid pool, without interfering with the lateral streams of high-velocity plasma gasses.

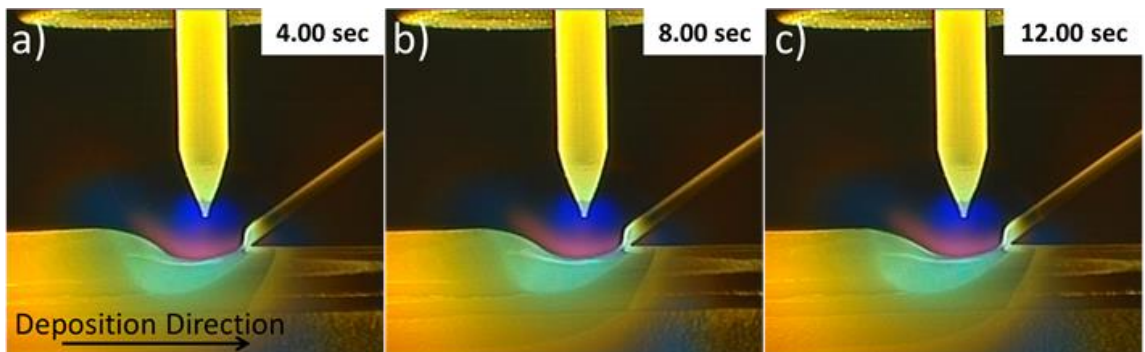

**Fig. 5:** Time-resolved images of the front-fed deposition process with a time frame of 4.0 seconds.

### 3.2   *Microstructure*

#### 3.2.1   Side-fed deposit

**Fig. 6a** shows the microstructural features of the structure deposited using the side wire feeding. The microstructure was composed of equiaxed grains near the bottom of the structure (**Fig. 6a** and **Fig. 6e**) and larger elongated grains on the upper part of the structure (**Fig. 6a** and **Fig. 6b**). The left-hand side (the wire feeding side) was characterised by a high presence of defects as lack of fusion, porosity and micro-cracks among the pores, as well as grain with a much finer grain size (**Fig. 6b** and **Fig. 6c**).

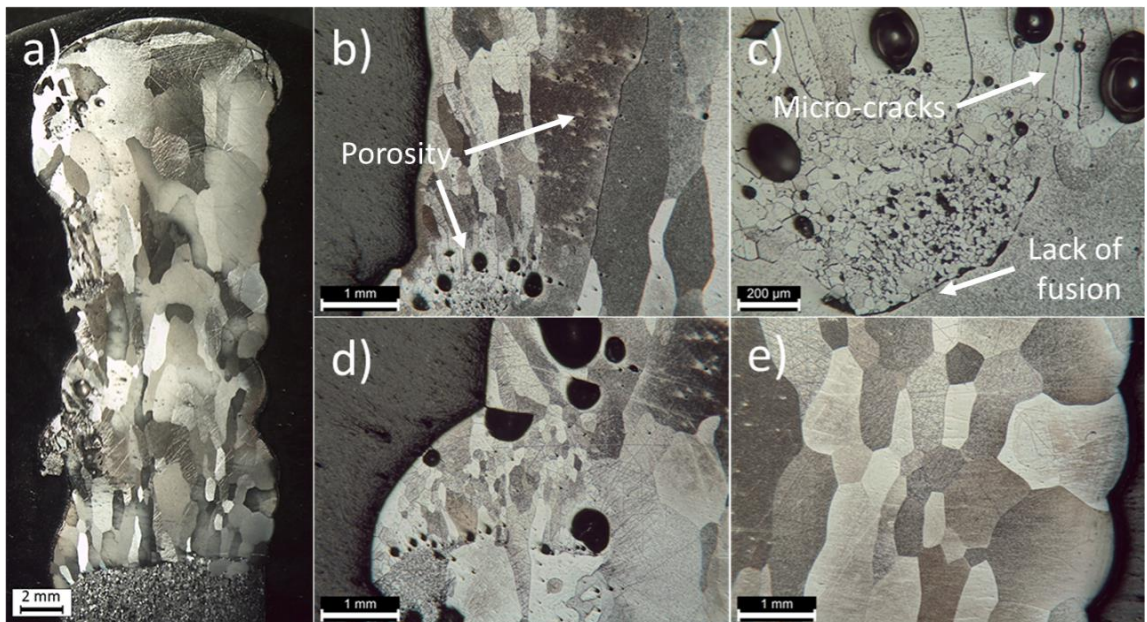

**Fig. 6**: Microstructure of the linear structure deposited using side wire feeding (a). Details of the defects located on the side facing the wire feeder (b-d). Detail of the microstructure at the base of the linear structure (e).

The formation of the equiaxed grains was promoted by the rapid solidification and the relatively high thermal conductivity at room temperature of tungsten [22]. For the deposition of the first layers, solidification nuclei started to form within the weld pool, favouring a multidirectional solidification from different nuclei during cooling. As the deposition progressed, the temperature gradually built-up leading to the reduction of thermal conductivity of tungsten [38–40] and therefore to a reduction of the thermal gradient. Under this condition, epitaxial grain growth occurred, resulting in elongated and coarser grain within the upper layers.

The grains around the defects on the wire feeding side were considerably smaller than in the rest of the structure, indicating that the wire feeding resulted in local cooling. The occurrence of porosity and lack of fusion was caused by the aforementioned particle ejections. The presence of these defects and the repetitive thermal cycles associated with the WAAM process led to the formation of a network of micro-cracks between the pores.

### 3.2.2 Front-fed deposit

**Fig. 7a** reports the microstructure of the structure produced using the front wire feeding. Similar to the side-fed structure, finer equiaxed grains (**Fig. 7a** and **Fig. 7e**) were found in the initial layers of the structure and coarser elongated grains (**Fig. 7a** and **Fig. 7c**) in the upper part of the deposit.

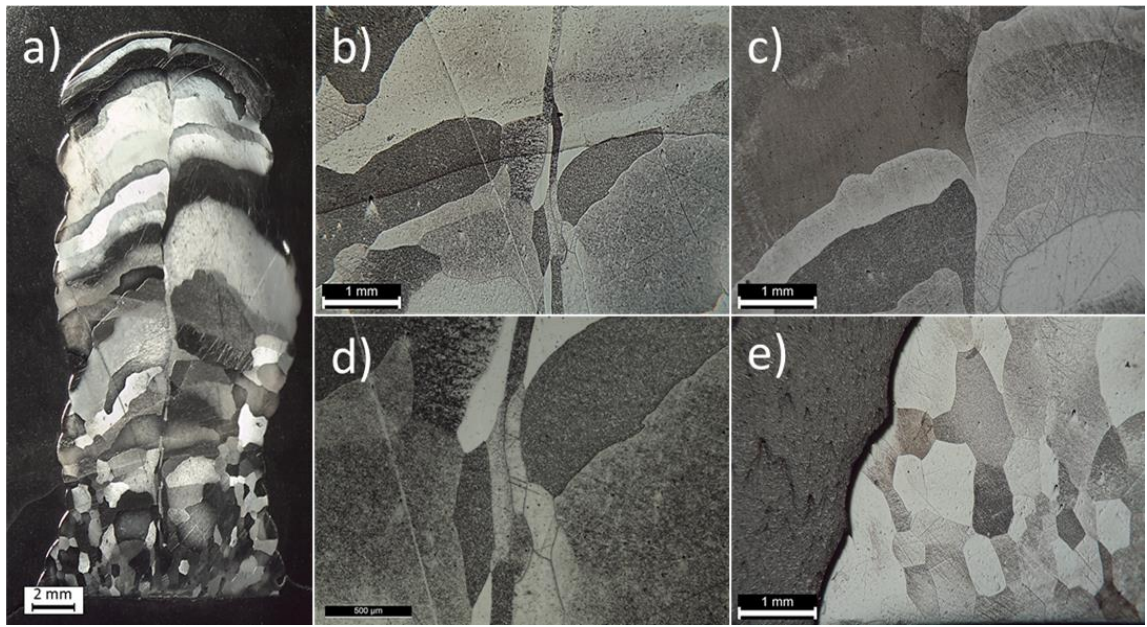

**Fig. 7**: Microstructure of the linear structure deposited using front wire feeding (a). Details of the elongated grains found at the centre of the structure (b, d). Detail of the microstructure of the upper part of the structure (c). Detail of the microstructure at the base of the linear structure (e).

The complete absence of pores and lack of fusion within the front-fed structure is directly correlated to the absence of spattered particles (**Fig. 5**). Clearly, avoiding spatter formation is required in order to preserve the structural integrity of the tungsten wall deposited via WAAM.

The grains of the upper part seemed to be the result of multiple specular dendritic structures propagating and meeting at the centre of the deposit (**Fig. 7c**).

### 3.3 *Deposition of large-scale tungsten wall*

The large-scale tungsten structure produced by WAAM, using the parameters reported in **Table 2** and the front wire feeding approach, is shown in **Fig. 8a** and **Fig. 8b**.

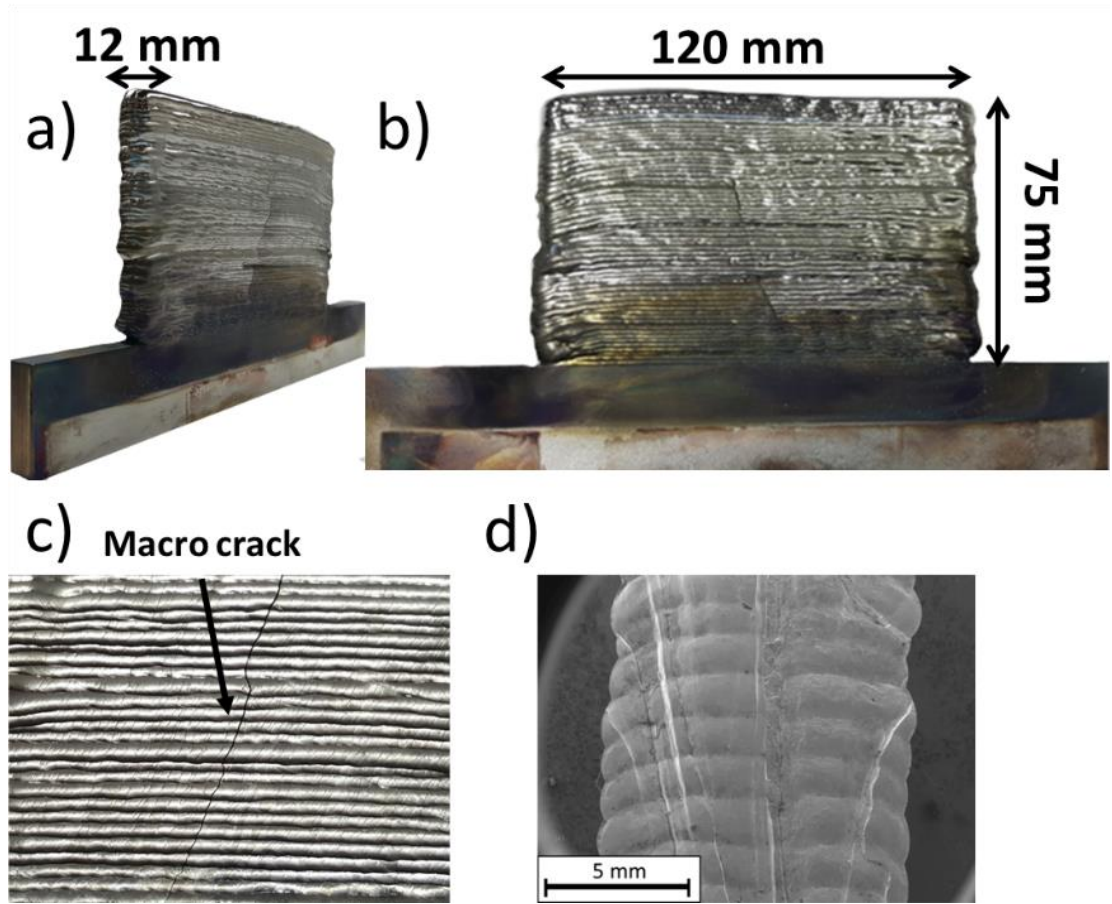

**Fig. 8**: Large-scale unalloyed tungsten linear structure deposited via Wire + Arc Additive Manufacturing process (a-b). Picture of the fracture from the outer surface (c) and SEM picture of the fractured surface (d).

This possibly represents the first fully-dense large-scale structure in unalloyed tungsten produced using AM. The unique aspects of this structure were the absence of any large network of grain boundary cracks within the volume deposited, the almost absence of discolouration and oxidation from the fusion process, and the consistency in the layers' geometry. **Fig. 8c** shows a detail of the surface of the linear structure which presents a macro-crack. The crack was characterised by an intergranular fractured surface and no additional intra-layer micro-cracks were observed (**Fig. 8d**). The occurrence of this single large crack was due to the deposition procedure. In particular, the upper half of the structure was deposited after leaving the structure to completely cool down to room temperature overnight. When the deposition process was restarted, a severe thermal shock developed within the already-deposited and highly-constrained region, giving origin to the crack. It stands to reason that this could be avoided by depositing without interruption, or by employing pre-heating prior to restarting the deposition.

## 4 Conclusion

In this study, it has been proven that the Wire + Arc Additive Manufacturing process is suitable for the production of large-scale refractory metal components by complete fusion. In particular, high-purity tungsten walls have been successfully deposited, without oxidation, and with high structural integrity and density. The main finding of this study can be summarised as:

- The orientation of the wire feeding has a severe impact on the occurrence of spatter and structural defects, such as pores and lack of fusion. In particular, a defect-free tungsten structure was deposited when using the front wire feeding approach;

- The microstructure of the deposit was also influenced by the orientation of wire feeding. A symmetrical and regular grain distribution has been observed for the structure deposited using front wire feeding, unlike, the structure produced using side wire feeding;

- The Wire + Arc additive Manufacturing technology could be considered a real candidate to replace the current method of manufacturing fully-dense large-scale components in tungsten.

## Acknowledgement

The authors wish to acknowledge financial support from the AMAZE Project, which was co-funded by the European Commission in the 7th Framework Programme (contract FP7-2012-NMP-ICT-FoF-313781), by the European Space Agency and by the individual partner organisations.

property of tungsten coatings prepared by vacuum plasma spraying technology, Fusion Eng. Des. 85 (2010) 1521–1526. doi:10.1016/j.fusengdes.2010.04.032.